\documentclass[10pt]{article}
\usepackage[T1]{fontenc}
\usepackage{graphicx}

\usepackage{hyperref}
\usepackage{color}

\usepackage{moreverb,url}

\usepackage{amssymb}

\usepackage{makecell}

\usepackage{xcolor} 

\usepackage{pifont} 

\usepackage[acronym,nomain,xindy]{glossaries} 
\setacronymstyle{long-sc-short}
\makeglossaries
\setacronymstyle{long-sc-short}

\newacronym{oaic}{OAIC}{Office of the Australian Informtion Commissioner}
\newacronym{gbp}{GBP}{Great British Pound}
\newacronym{asbfeo}{ASBFEO}{Australian Small Business and Family Enterprise Ombudsman}
\newacronym{asd}{ASD}{Australian Signals Directorate}
\newacronym{nist}{NIST}{National Institute of Standards and Technology}
\newacronym{ncsc}{NCSC}{National Cyber Security Centre}
\newacronym{acsc}{ACSC}{Australian Cyber Security Center}
\newacronym{abn}{ABN}{Australian Business Number}
\newacronym{fte}{FTE}{Full Time Equivalent}
\newacronym{osint}{OSINT}{Open Source Intelligence}
\newacronym{asic}{ASIC}{Australian Securities \& Investment Commission}
\newacronym{sme}{SME}{Small \& Medium Enterprise}
\newacronym{iaas}{IaaS}{Infrastructure as a Service}
\newacronym{paas}{PaaS}{Platform as a Service}
\newacronym{saas}{SaaS}{Software as a Service}
\newacronym{ablis}{ABLIS}{Australian Business License and Information System}
\newacronym{abs}{ABS}{Australian Bureau of Statistics}
\newacronym{pmt}{PMT}{Protection Motivation Theory}
\newacronym{soc}{SOC}{Stages of Change}
\newacronym{raa}{RAA}{Reasoned Action Approach}
\newacronym{sitd}{SITD}{Small IT Data}
\newacronym{iot}{IoT}{Internet of Things}
\newacronym{pcidss}{PCI-DSS}{Payment Card Industry Data Security Standard}
\newacronym{mfa}{MFA}{Multi-Factor Authentication}
\newacronym{ciso}{CISO}{Chief Information Security Officer}
\newacronym{hia}{HIA}{Housing Industry Association}
\newacronym{mba}{MBA}{Master Builders Association}
\newacronym{bim}{BIM}{Building Information Modeling}
\newacronym{oss}{OSS}{Operations Support System}
\newacronym{afma}{AFMA}{Australian Fisheries Management Authority}

\begin{document}
\title{Structuring the Chaos: Enabling Small Business Cyber-Security Risks \& Assets Modelling with a UML Class Model}

\date{}

\author{
{\rm Tracy Tam}\\
RMIT University, Melbourne, Victoria
\and
{\rm Asha Rao}\\
RMIT University, Melbourne, Victoria
\and
{\rm Joanne Hall}\\
RMIT University, Melbourne, Victoria
}

\maketitle              
%
\begin{abstract}
Small businesses around the world are increasingly adopting IT,  and consequently becoming more vulnerable to cyber-incidents.  Whilst small businesses are aware of the cyber-security risks around their business, many struggle with implementing mitigations. Some of these can be traced to fundamental differences in the characteristics of small business versus large enterprises where modern cyber-security solutions are widely deployed. 

Small business specific cyber-security tools are needed. Currently available cyber-security tools and standards assume technical expertise and time resources often not practical for small businesses.  Cyber-security competes with other roles that small business owners take on, e.g. cleaning, sales etc. A small business model, salient and implementable at-scale, with simplified non-specialist terminologies and presentation is needed to encourage sustained participation of all stakeholders, not just technical ones. 

We propose a new UML class (\gls{sitd}) model to support the often chaotic information-gathering phase of a small business' first foray into cyber-security.  The \gls{sitd} model is designed in the UML format to help small business implement technical solutions. The \gls{sitd} model structure stays relevant by using generic classes and structures that evolve with technology and environmental changes. The \gls{sitd} model keeps security decisions proportionate to the business by highlighting relationships between business strategy tasks and IT infrastructure.  

We construct a set of design principles to address small business cyber-security needs.  Model components are designed in response to these needs. The uses of the \gls{sitd} model are then demonstrated and design principles validated by examining a case study of a real small business operational and IT information.   The \gls{sitd} model's ability to illustrate breach information is also demonstrated using the NotPetya incident. 

\end{abstract}
%
%
%
\section{Introduction}
Small business (0-19 employees~\cite{AustralianBureauofStatistics2001}) plays a crucial role in the global economy, measured by the number of small enterprises~\cite{DepartmentforBusinessEnergy&IndustrialStrategy2021}, employment statistics~\cite{ASBFEO2020,OECD1997} and other contributions~\cite{Park2018,Sullivan-Taylor2011}. The pandemic propelled small businesses globally into a new world driven by technology. Small business cyber-security has, thus, become a problem that can no longer be ignored.  As the number of novice small businesses adopting technology increases~\cite{Chiappetta2020}, so does the cyber attack surface~\cite{Tam2020}.  Getting small businesses to a cyber-security ready state has proved challenging, with many small business owners aware of the need for cyber-security but unsure of what to do~\cite{AustralianCyberSecurityCentre2020a,SmallBusinessDigitalTaskforce2018}. 

Small business differs from larger enterprises in many ways, ranging from communication style, financial resources, internal expertise available, to a lack of immediate incentives ~\cite{Tam2021}. Unlike larger enterprises, the majority of small businesses have less time and resources to study and manage IT and cyber-security.  Small businesses with less than 5 employees make up nearly 90\% of all businesses in Australia \cite{AustralianSmallBusinessandFamilyEnterpriseOmbudsman2023}. In these micro businesses, a business owner often takes on ancillary jobs such as cleaner, security guard and IT support, in addition to core business responsibilities. Cyber-security is only one of many competing priorities (e.g. generating sales, making products etc.) needed to keep the business alive. 

Cyber-security analysis of a small business is often left up to individual implementers, using tools meant for large businesses, producing sub-optimal results for this resource-constrained cohort. \cite{Tam2021}. There is need for an effective model to organise and analyse critical small business cyber-security information. 

Cyber-security tools and processes lack focus on small business priorities and are not proportionate to the size of a small business. Fundamental to any new tools or processes is a consistent and enduring way of organising cyber-security information for subsequent analysis. Here, inspiration could come from IT where sorting and storing of information is often done via data models \cite{Microsoft2023}. A data model foundation allows rapid deployment in various solution technology stacks for any potential small business security solutions. 

This paper proposes a UML data model, the \acrfull{sitd} model, and process to support the often overwhelming information-gathering phase of a small business cyber-security journey. We first describe the rationale behind the new approach and model in section~\ref{sec:ProblemExistingTools}. Section~\ref{sec:ModelDeisgnPrinciples} describes the design principles. In section~\ref{sec:ModelToolChoice} we justify the choice of UML as a modelling tool. The new class model and its components are introduced in section~\ref{sec:TheModel}, before validation via application to real-life small businesses (case study) in section~\ref{sec:ApplicationModel}.  The insights and analysis gained from the resulting models are discussed throughout. Section~\ref{sec:Discussion} discusses how the model's aligns with the aims and design principles.

\section{Problems with Existing Tools}
\label{sec:ProblemExistingTools}
Many existing cyber-security analyses and tools are designed with the flexibility to allow for use in any organisation.  This flexibility requires inherent technical expertise to understand and apply appropriately. This expertise is not available within the small business space ~\cite{Tam2021}.

The following issues arise when a small business tries to use current tools:
\begin{itemize}
\item \textbf{Attack Based Lists} (e.g. OWASP \cite{Watson2018}, Mitre Att\&ck \cite{TheMitreCorporation2019}) -- A business needs to navigate through a repository of attack techniques with no relation to business priorities.  IT training is needed to understand the terminology used, as well as the implications and issues discussed.
\item \textbf{Controls Based Lists} (e.g. CIS Controls \cite{CenterforInternetSecurity2023}, Essential 8 \cite{AustralianCyberSecurityCentre2021a}) -- Use of generic terms e.g. systems, applications, requires a small business to further investigate what the controls apply to.  This risks things, e.g. IoT devices, being left out altogether because they don't fall into an easily recognisable category .
\item \textbf{Risk Management Based Standards} (e.g. ISO 27001 \cite{ISO27001}, NIST Cybersecurity Framework \cite{NIST2018}) -- Include broad statements that require interpretation of how controls and processes apply to a business.  The results of this process often have technical consequences which are not obvious without technical training.
\end{itemize}
The vast majority of small businesses have no technical expertise~\cite{AustralianBureauofStatistics2021}. With limited turnover and budget, the average small business cannot afford a cyber security professional~\cite{Tam2021}.  As a non-revenue-generating task, cyber-security needs to be understandable and usable by these non-technical small business decision makers. A more approachable tool, from both technical and investment perspectives, is needed to make cyber-security accessible for small businesses, making participation more attractive. 

\subsection{New Approach Needed}
\label{subsec:newapproachneeded}

The lack of tailoring to small businesses undermines the effectiveness of any analytic tools or assistance.  New approaches are needed to take into account the human aspects of small businesses~\cite{Renaud2016,Bada2019a,Tam2021}. Humans (employees and owner) are a critical part of cyber-security plans and posture~\cite{VonSolms2013}. In addition, a good security strategy requires involvement from stakeholders at all levels in a business ~\cite{Tan2017}. 

Maintaining cyber-security ownership over the long term in a small business can be challenging when many cyber-security resources require a level of technical understanding. The majority of adults in developed countries do not have a high level of technical skills~\cite{OECD2018a}. Expecting non-technical small business owners to gain the requisite cyber-security technical knowledge unassisted undermines a business' cyber-security self-efficacy. Lack of self-belief in the effectiveness of one's action can undermine the motivation to engage~\cite{Maddux1983}. According to the EAST principle~\cite{Service2014} of human behavioural insights, making something easy also helps encourage the desired behaviour. Improving the overall ease of access for non-technical stakeholders can improve the outcome of a cyber-security solution by facilitating effective and ongoing participation.

To enable small business engagement with any new cyber-security tool, it is essential to redesign the current highly technical mental structure of cyber-security analysis. A cyber-security system with infinite flexibility has proven to be an issue for small business adoption (as mentioned above), so any new approach has to address this gap by tailoring to small business operational and human characteristics. To achieve this, a re-defining of the fundamental building blocks of how cyber-security is presented to small business is needed. Technology and cyber-risk-centric language, common in existing security frameworks, need to give way to concepts familiar to non-technical business owners.  The proposed data model needs to center on business concepts and terms that an average person with no technical training can understand. 

Building on the above, to support a business-friendly cyber-security process, we propose a new class model (referred to as \gls{sitd} model) to help a small business organise and record its priorities and its IT. For a small business operator, a model needs to document IT details as these are often not at the forefront of their mind \cite{Scott1987}, with many relying on informal communication \cite{Street2004}.  The initial cyber-security step of recording the IT being used can require significant effort from small businesses. The \gls{sitd} model eases the cognitive load by beginning with information in areas of familiarity, e.g. day-to-day tasks, before tying it back to cyber-security.

\section{Target Users and Businesses of \gls{sitd} Model}
\label{sec:TargetUsersBusinesses}
At a technical level, the \gls{sitd} model serves as the data foundation for future small business cyber-security analysis tools and processes. There are 2 primary target stakeholders for the \gls{sitd} model: the small businesses to be protected, and the cyber-professionals implementing cyber-security tools to be used by the small business.

\subsection{Target Modelled Businesses}
The \gls{sitd} model is designed to model small businesses with 0 -19 employees \cite{AustralianBureauofStatistics2001}, in order to protect them. (Sole traders/single person companies are considered to have 0 employees.) One exclusion to this broad scope is small businesses that offer IT-centric services and products, as they are likely to have additional security considerations (e.g. DevSecOps \cite{Microsoft2023a}) and technical skills.

\subsection{Target Data Structure Users}
\label{subsec:TargetDataStructureUsers}
The \gls{sitd} model is created for modelling non-technical small businesses. The large number of small businesses means that any small business cyber-security tools need to be deployable at-scale. Technology will likely play a vital role in the tool's dissemination. Consequently, the \gls{sitd} model needs to provide a structure that is easily implementable using technology. Due to this, the more immediate users of the \gls{sitd} data model are likely to be prospective cyber-security tool developers or professionals looking for a common way to structure information in a quest to protect small businesses. 

The \gls{sitd} model gives developers and professionals the data foundation for future security tools with a small business-centric approach. The model will help solution implementers to move away from the traditional cyber-security and technology centric approaches.

\section{Model Design Principles}
\label{sec:ModelDeisgnPrinciples}
In addition to the small business and developer characteristics discussed in Section \ref{sec:TargetUsersBusinesses}, the \gls{sitd} model is designed with the following guiding principles to facilitate small business applications.

\subsection{Focus on Business Priorities}
\gls{sitd} model needs to start from the perspective of business goals, the job tasks supporting these goals, and the IT tools needed to support these job tasks.  The goal of cyber-security is to facilitate secure business use of IT.  To be used, IT needs to deliver value to the business.  In small businesses, IT tools are often chosen because the IT solution is more convenient or efficient than its manual counterpart (e.g. digital spreadsheet for sales recording). 

The re-focus towards business priorities also serves as a reminder that the business value of the IT tool needs to be preserved, even after securing it. However, business value cannot be judged from a technology-centric view alone.  Another important consideration is ensuring incident response and recovery are business goals.  

\subsection{Capture Intangible Factors and Relationships}
The \gls{sitd} model needs to illustrate relationships between physical and intangible components. 

Past cyber-security incidents and attack techniques have involved many factors ranging from technology and operations to people \cite{Greenberg2018,Greenberg2021}. Social engineering attacks illustrate the need for technology to work with the less tangible aspects of an organisation such as human behaviour. The \gls{sitd} model needs to capture the impacts of the non-tangible factors on the overall posture of a business, e.g. being in a highly competitive field (industrial espionage), or having disgruntled employees.

The documentation of physical and intangible factors provides opportunities for using the defence in depth security strategy \cite{NationalInstituteofStandardsandTechnology2020}.  This strategy allows smaller tools/solutions to be combined to form better overall security postures. The ability to combine smaller controls is advantageous to small business due to their resource constraints \cite{Tam2021}, which makes them unlikely to be able to afford large ready-made cyber-security solutions.

\subsection{Allow for Incomplete Information}
The threat landscape is ever-evolving~\cite{Ande2020,Colbaugh2011} and cyber-security is never `finalised'. Hence, the \gls{sitd} model needs to support cyber-security as a continuous improvement process along with the need for it to function even in an incomplete state.   In addition, businesses, particularly those in a startup phase (especially small businesses) often change on a day to day basis.  As such, tools expecting linear progression or completion (e.g. waterfall methodologies) are not practicable from either a logistics or ongoing relevance perspective. Therefore, the \gls{sitd} model needs to view incomplete information as an opportunity for further discussion and exploration, rather than a roadblock.  

\subsection{Agnostic}
For maintainability into the future the \gls{sitd} model should be agnostic with regards to several factors:
\begin{enumerate}
\item \textbf{Language} -- Cyber-security crosses international borders: many cyber-crimes have transnational elements~\cite{UnitedNationsOfficeonDrugsandCrime2013}. Minimising textual information provides room for visual representations. A combination of textual and visual information has been shown to increase understanding~\cite{Angeli2004} and recall~\cite{Blanco2010}. Where possible the \gls{sitd}  model needs to prioritise visual cues over words.
\item \textbf{Technology} -- Where applicable the \gls{sitd} model needs to be technology agnostic to ensure continued relevance as technology or attack techniques evolve.  The \gls{sitd} model needs to serve as a record of the ongoing relevance of IT to the small business, and not as a static picture.
\item \textbf{Standards, Legislation and Industry} -- A mobile phone used in a food service business essentially has the same technical vulnerabilities as the same one in a legal practice.  However, the context of supported business goals sets these 2 phones apart as do the ramifications if the phenes are breached or lost.  Rather than specifically including locale-specific conditions such as HIPAA, GDPR, Reportable Breach scheme etc, the \gls{sitd} model needs to capture local context in the business part of the model, e.g. job function, tasks, strategies.
\end{enumerate}

\subsection{Enable Cyber-Security Analysis}
Ultimately the \gls{sitd} model should aim to capture sufficient information about the small business to enable further risk analysis and discussions. Most cyber-security standards~\cite{NationalCyberSecurityCentre2021} emphasise the need for IT inventory recording.  The \gls{sitd} record, especially when linked to its users, needs to serves as input to allow planning of further security steps for the business.

In addition, the \gls{sitd} model needs to capture sufficient detail to conduct cyber-security risk analysis, including assessing Common Vulnerability and Exposures (CVE) impacts, risk mitigation planning and incident analysis.

\section{Choice of Modelling Tool}
\label{sec:ModelToolChoice}
Our choice of the \gls{sitd} model tool is driven by the needs of potential implementers of small business cyber-security solutions - IT developers and cyber-security professionals, for reasons discussed in Section \ref{subsec:TargetDataStructureUsers}. 

Hence we focus on data model formats common within the IT industry. The following candidate approaches were identified and evaluated:
\begin{itemize}
\item Spreadsheets/Data Tables with Custom Relationships. 
\item Entity Relationship Diagram (ERD)~\cite{Chen1976}.
\item Unified Modelling Language (UML)~\cite{InternationalOrganizationforStandardization2005}.
\end{itemize}

The table/databases approach is unsuitable due to lack of a standardised way for entities and relationships to be modelled in cyber-security.  The table approach requires creation of a new ecosystem: with new rules needing to be created, tested and maintained.  A new standard only serves to add complexity within a field already overflowing with information.

\subsection{Advantages of UML}
\label{subsec:advantagesofUML}
While small business IT can technically be modelled within both ERD and UML, UML is more suitable for the following reasons:
\begin{itemize}
\item \textbf{Holistic nature of the UML ecosystem} -- UML encompasses technical and non-technical aspects.  This aligns with the aim of describing structural, intangible and dynamic aspects of small business cyber-security. For future work, UML can inherently model behaviour, business processes, timing, actions etc.
\item \textbf{UML allows for different perspectives} -- UML natively recognises that the same piece of data can be viewed from different perspectives (or `views').  For example, customer information is a piece of data inside a business database, but UML can recognise that it is used by staff when a customer enquires about their account. Given that the \gls{sitd} model is concerned with cyber-security in both the data itself as well as transitory events on the data, UML's view better aligns with our aims.
\item \textbf{UML can be converted into ERD} -- A UML class diagram can be converted to ERD, making, reversion to ERD possible in the future.
\item \textbf{Slimline presentation of information} -- UML allows for attributes and operations to be applied to individual entities. A similar entity within ERD requires multiple objects making the entity more complex in visual appearance.
\item \textbf{Use of UML within the industry} -- UML has been adopted widely into the IT and business worlds, where it can be used to describe whole ecosystems of software, hardware, processes and actors involved.  This focus is important given the potential developers of future small business security solutions.
\end{itemize}

In conclusion, UML design philosophy is better aligned with the \gls{sitd} model goals and the possible future extensibility of small business cyber-security. 
\section{The \gls{sitd} Model}
\label{sec:TheModel}
In this section, we present details of each sub-part of the \gls{sitd} model, before linking the parts together into an overall structure.  Real life business operations and IT architectures are then used to illustrate the model's use. The \gls{sitd} model centers on the business and branches out to the connected IT infrastructure.

The \gls{sitd} model relies heavily on the class and object diagrams within UML. These diagrams operate on the concept of class and associations between these classes.  All other details are defined within this construct. For a quick start guide on reading UML models and conventions, please see Appendix ~\ref{sec:AppendixReadUML}.

\subsection{The \gls{sitd} Model Classes}
\label{subsec:ClassAreas}
The classes in the \gls{sitd} model reflect important small business and cyber-security entities, and as such are not restricted to physical entities.  Table \ref{tab:ClassAreas} identifies critical concepts which are treated as classes in the \gls{sitd} model.

\begin{table}[h!]
\begin{tabular}{|p{5.5cm}|p{6.5cm}|}
\hline
\textbf{Class Areas}&\textbf{Examples}\\
\hline
Business entity & Doe's Gardening Service \\
\hline
Human/group actors & Owner; Employees\\
\hline
Job tasks \& roles & Payroll Processing; Manufacturing\\
\hline
Physical location & Office; Client Site\\
\hline
Selected hardware \& software details & Mobile Phone; CAD\\
\hline
Remote/cloud IT systems & Client IT Systems; Cloud Data Storage\\
\hline
Data/information & Customer Data; Audit Records\\
\hline
Motivations and strategies & Stay in Business; Lifestyle; Product Quality\\
\hline
\end{tabular}
\caption{\label{tab:ClassAreas} Class areas (discussed in section~\ref{subsec:ClassAreas}) covered as part of the \gls{sitd} model's aim of capturing cyber and business salient information.}
\end{table}

Note: Some concepts, described in multiple classes within IT architecture design and modelling conventions, have been simplified into a single construct. The reasons and implications are discussed in each part below.
Associations (used to describe class relationships between 2 classes) contain additional information, including relationship multiplicity as well as direction.  These are noted in standard UML notations. 

In the next four subsections, we will look at sub-components of the \gls{sitd} model. Section~\ref{sec:OverallModel} connects the sub-models back into the overall model structure. 

\subsection{\gls{sitd} Submodel: Business}
\label{sec:BusinessModel}
\begin{figure}[h]
{\centering
\includegraphics[width=\textwidth]{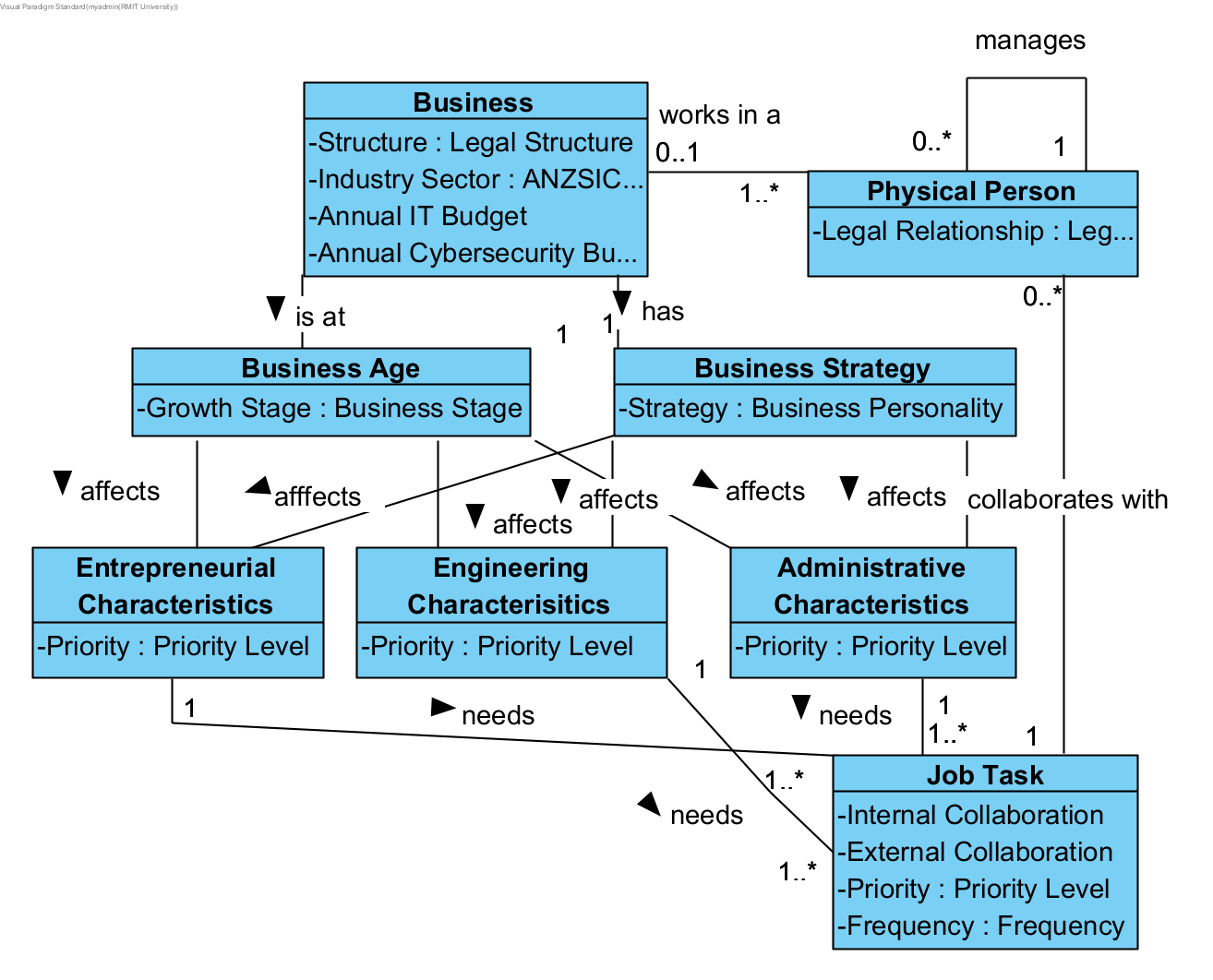}
\caption{\label{fig:BusinessStrategy} UML class structure diagram showing the connections between different parts of the business. (Described in section~\ref{sec:BusinessModel}.)}
}
\end{figure}

To ensure security decisions are relevant and proportional, the business is central to the \gls{sitd} model.  The \textit{Business Strategy} of the \gls{sitd} model (Figure~\ref{fig:BusinessStrategy}) captures the fundamental relationships between the business, its aims and the people working in it. 

The base class is centred on the business entity being protected, i.e. the business itself. Basic details of the business are captured within the attributes. The business' maturity stage and strategy~\cite{Miles1978}, which affects its behaviour profile~\cite{Scott1987}, are recorded for influence on business activities.  The derived characteristics are placed into 3 broad categories, \textit{Entrepreneurial, Administrative} and \textit{Engineering}, which are then linked to individual day-to-day tasks. Entrepreneurial characteristics capture any growth ambitions e.g. partnerships, branding, while Engineering characteristics describe operational matters e.g. factory building.  Administrative characteristics keep the business running, e.g. business registration, tax management, legal compliance. (For businesses with narrow business goals, the characteristics and job tasks can be combined into a single layer to streamline the \gls{sitd} model.) 

This view of the \gls{sitd} model highlights that without the business reasons stemming from the business strategy (e.g. keeping the business alive), the tasks would not be done.

The number of physical persons (and management hierarchy) working within a business is also recorded, since business is a human endeavour.  In a small business, the distinction of physical people is important given the manual (and often adhoc) processes that exist \cite{Miles1978}. The multiple hats worn by small business employees leads to informal and less defined communication and processes than those in larger corporations.  Since human communication forms the basis of many social engineering attacks~\cite{Mouton2014}, the number of employees is an important part of cyber-security strategies.

\subsection{\gls{sitd} Submodel: Job Function}
\label{sec:JobFunctionModel}
\begin{figure}
{\centering
\includegraphics[width=\textwidth]{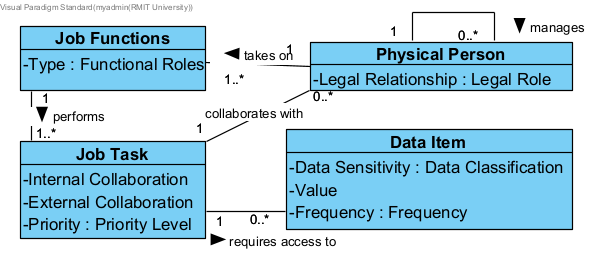}
\caption{\label{fig:JobFunction} The UML class structure diagram showing how job functions lead to people working together and hence needing access to specific data within the business. (Details in section~\ref{sec:JobFunctionModel}.)}
}
\end{figure}

The \textit{Job Function} part of the \gls{sitd} model describes the links between job tasks and the roles performing the tasks (Figure ~\ref{fig:JobFunction}).

The job tasks described as part of business goals are linked to the human responsible for the task.  The task is performed in the context of the person's job role.  The \gls{sitd} model allows for more than one person to work on the same task using a collaboration link in the context of different roles (e.g. business plans being completed in collaboration between the owner and accountant).  Conversely, multiple physical people can work on the same job task via the same function roles (e.g. multiple sales assistants in a retail shop). This model allows for the same physical person to take on multiple job functions, as is common in micro-enterprises where the owner can also be the security, janitor and website administrator. In early analysis, the function role can be synonymous with a physical person; role information can be added after further analysis.  This is particularly useful in sole trader/micro-companies where roles are not formally assigned or defined.

The job function model links job tasks to the data item(s) needed to perform the tasks.  Every business task requires some sort of data item to be used, whether it is the product price in a sales transaction, a recipe within a manufacturing plant, or lesson plans within an education setting.  Data is considered at a more fundamental level than just the electronic storage of the information.
										
\subsection{\gls{sitd} Submodel: IT Interaction}
\label{sec:ITInteractionModel}
\begin{figure*}
{\centering
\includegraphics[width=\textwidth]{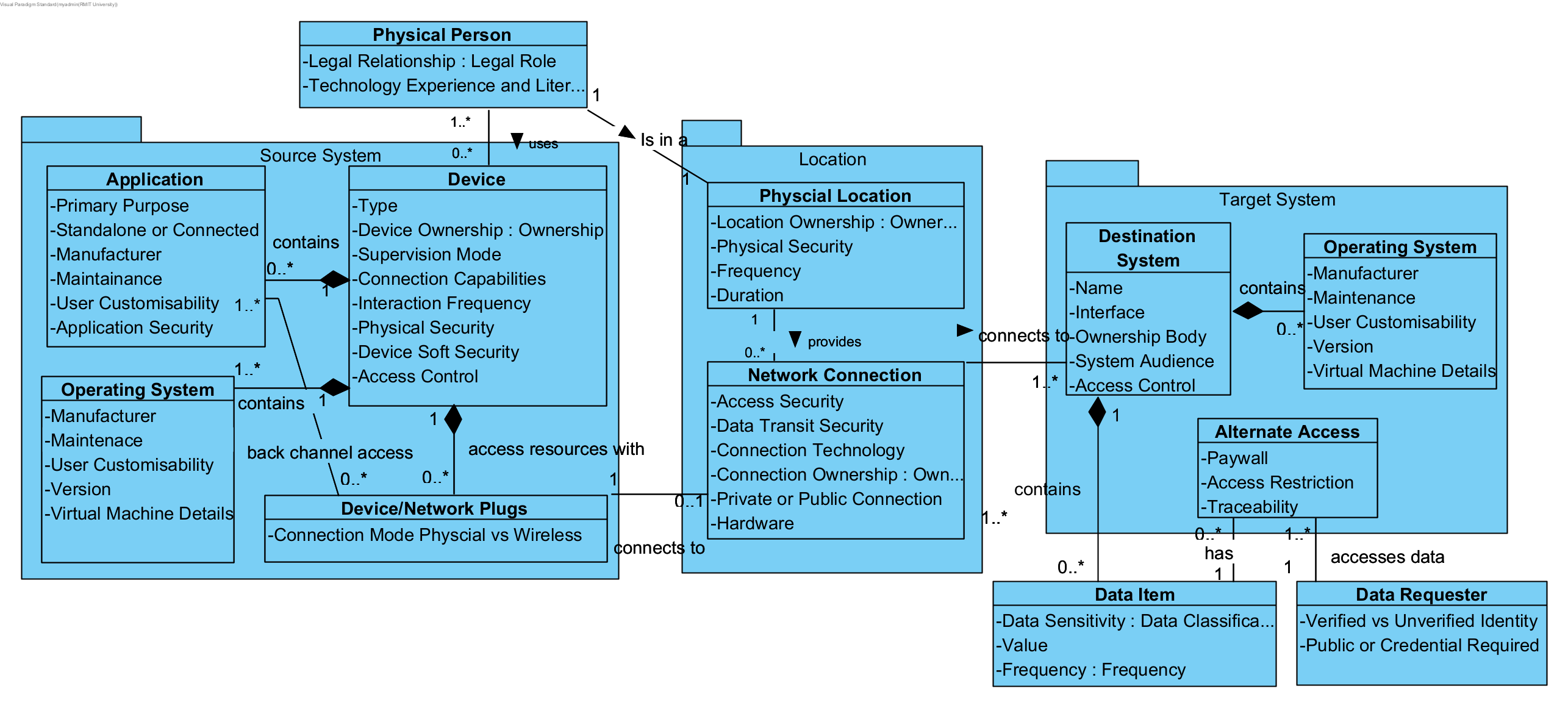}
\caption{\label{fig:ITAccess} UML class structure diagram showing how a person accesses specific data in target systems using their devices and network connection.   (Discussed in section~\ref{sec:ITInteractionModel}.)}
}
\end{figure*}

We now link data to the IT systems used to store and use this data (Figure~\ref{fig:ITAccess}). The \gls{sitd} model only includes any data stored electronically.  It is assumed that any physical data stored is handled according to the existing risk management plan and outside the scope of cyber-security. Inspite of the \gls{sitd} model's electronic focus, ISO27001 \cite{ISO27001} includes physical access as part of risk management. Hence, any system implementing the \gls{sitd} model needs to include reminders that physical security is still required.

The data item must be stored within a destination (target) system.  This is intended to be a generic container that records details of where the data is stored.  The target system can range from a local drive on a laptop/phone to cloud services or records held by another business/entity. The details of the service and location are captured in the classes associated with the data. We recognise that data (class Alternate Access) can be retrieved by another party off the same system, e.g. business registration details can be requested by a member of the public using the registration body website.  The target system classes are deliberately light on technical detail to reflect the reality that most small businesses have limited influence on the technical details of the electronic data storage.  For example, hosted websites only allow limited customisation from a look and feel perspective, webmail providers dictate the login process and whether multi-factor authentication mechanisms are offered, retail hardware/software manufacturers decide whether memory is encrypted by default etc.

On a physical front, the \gls{sitd} model links the access to the data via a physical location and device, and ultimately to a physical person. The link highlights that physical access to this device-connection pipeline plays an important part in the security posture.  For example, if someone has physical access to a Wi-Fi router, then no matter how secure the target system, the risk of a man in the middle attack increases.

At a physical device level, the classes are simplified from a technical viewpoint e.g. the OSI 7 layer model \cite{Day1983}, to capture only applications, operating system (OS) and network connector classes - components the business worker interacts with.  This is to ensure focus on the components under the worker's control on a day to day basis. An application is any software program that the business uses on a physical device, ranging from productivity suites such as Microsoft Office to browsers for access to cloud services/web pages. While the simplification does make technical vulnerability analysis more difficult, most vulnerability notifications, common vulnerabilities \& exploits, vendor notifications today relate impacts to applications and/or operating systems \cite{TheMitreCorporation2023}.

In its current version, the \gls{sitd} model does not consider system to system IT events e.g. batch jobs, scheduled events. The \gls{sitd} model's target audience are non-technical small businesses; the utilisation of automated events is minimal~\cite{AustralianBureauofStatistics8167}.

\subsection{\gls{sitd} Submodel: Threats}
\label{sec:ThreatModel}
Finally, to illustrate deviations in risk between different industries and businesses, a threat model is included in the \gls{sitd} model to describe any specific or general threat to the business.  This section again focuses on human threat actors rather than the technical threat. Removing human motivation eliminates many reasons for exploiting a vulnerable system.  This human threat actor class can describe single actors e.g. an industry competitor, or groups  e.g. Advanced Persistent Threat (APT)/nation state actors.

\begin{figure}[h]
{\centering
\includegraphics[width=\textwidth]{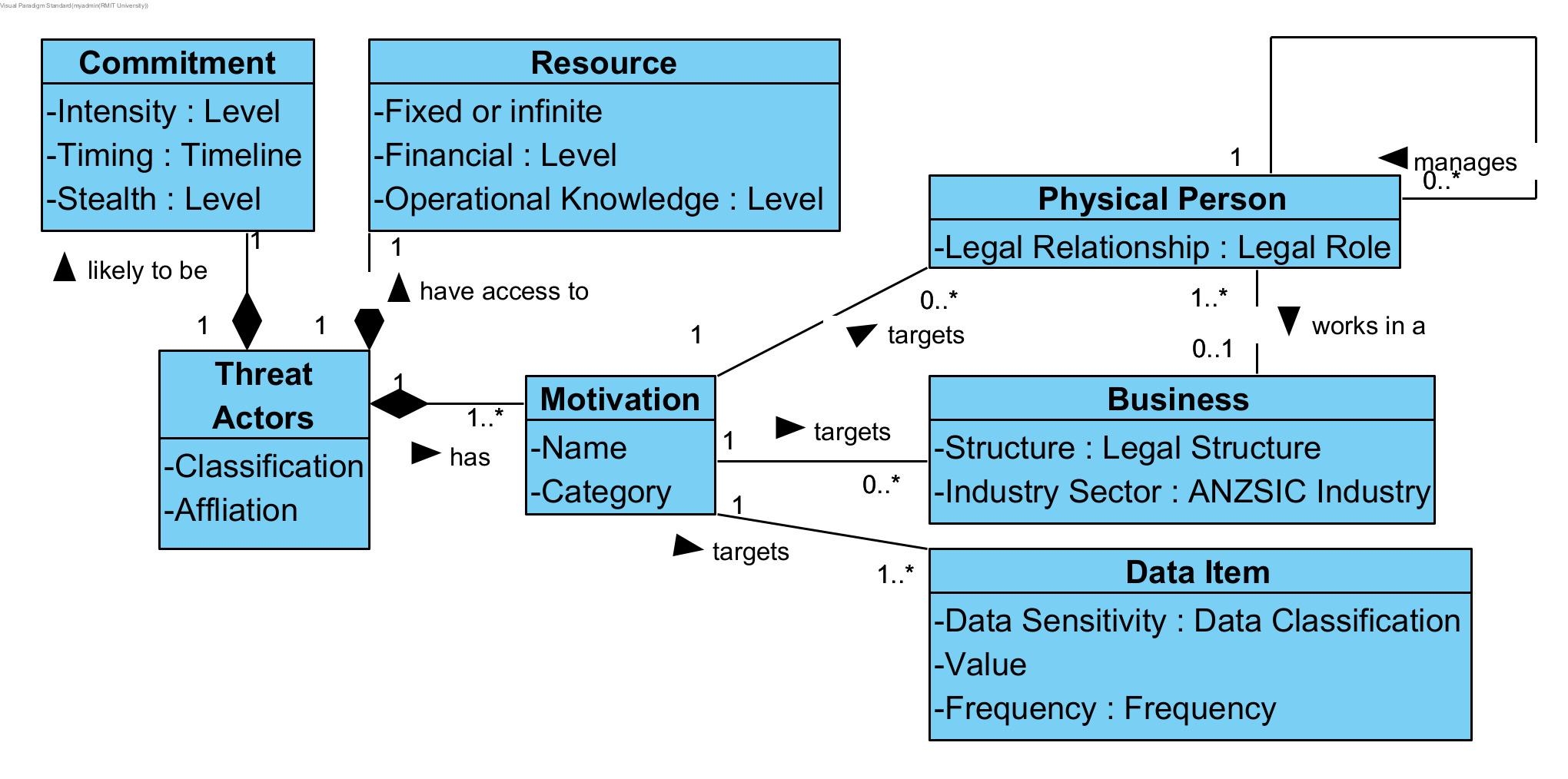}
\caption{\label{fig:ThreatLandscape} UML class structure diagram showing how certain threat motivations can mean specific data (and IT system by association) can require more attention. (Discussed in section~\ref{sec:ThreatModel}.)}
}
\end{figure}

The majority of small businesses, before having a cyber-incident, do not perceive cyber incidents as likely~\cite{AustralianCyberSecurityCentre2020a}. The threat model is important to document motivations, especially in specialised industries (e.g. defence contractors, fiercely competitive market conditions etc.). The threat model highlights any part of the business that may require a higher level of priority, to help assess adequate level of investment.

\subsection{Overall \gls{sitd} Model}
\label{sec:OverallModel}
\begin{figure*}[h]
{\centering
\includegraphics[width=\textwidth]{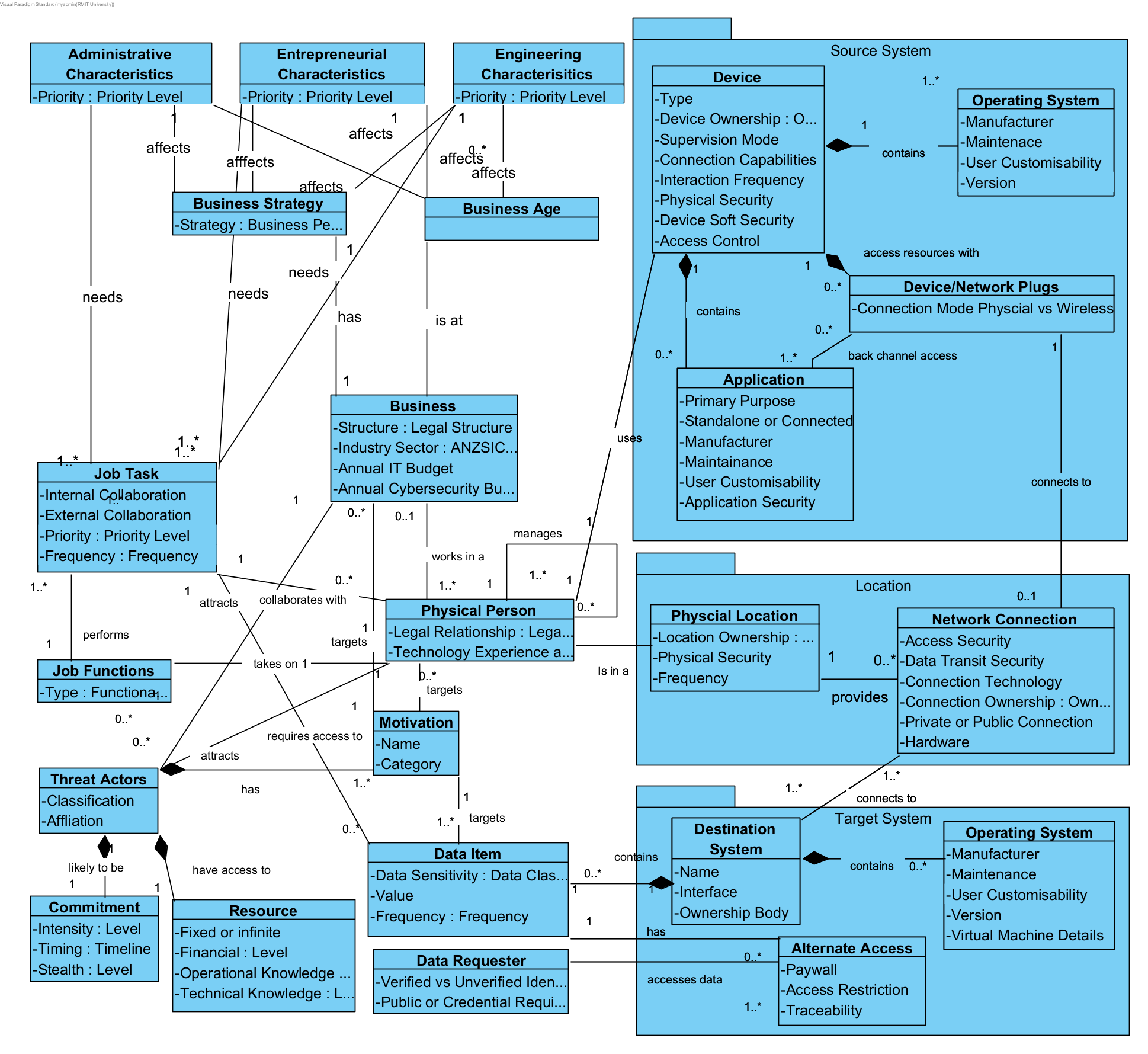}
\caption{\label{fig:ClassOverview} UML class overview showing the relationships between various factors within a (sample) small businesses' overall cyber-security posture, discussed in section~\ref{sec:OverallModel}.}
}
\end{figure*}

Each subpart of the \gls{sitd} model described in sections~\ref{sec:BusinessModel} -~\ref{sec:ThreatModel} can be used independently. However, when linked together with common classes, the model creates a picture of the interrelations between IT and business goals (Figure~\ref{fig:ClassOverview}).  Based on the linked business goal, the business can prioritise the parts of IT needing attention from a cyber-security perspective.

From an analytics perspective, the relationships show how unconnected parts of the business can lead to an asset needing protection from motivated actors.

A class and relationship model highlights gaps without impeding progress. Partial data is of use to trigger investigation and further discussions.

\section{\gls{sitd} Model Applications}
\label{sec:ApplicationModel}
We will now demonstrate the use of the \gls{sitd} model by modelling a case study business. Business operation modelling of cyber-security relevant concerns is done using source data from a non-cyber-security small business case study \cite{L2005} \footnote{As the business is currently operating, all identifying information of the small business, individuals, specific products and locations are redacted in the public version of this article to protect their privacy, including citations. Original information and citations were provided for review purposes.} to emulate a small business owner's point of view. Technical modelling is sourced from a UK small scale IT architecture case study \cite{Osborn2017} , including small business participants. Finally, we utilise a NotPetya breach analysis to illustrate the \gls{sitd} model use in incident analysis.

\subsection{Modelling Business Point of View}
\label{subsec:ModelBusinessPoV}
An academic case study of an agricultural small business \cite{L2005} is used to demonstrate the \gls{sitd} model.  The case study was originally constructed to illustrate quality control considerations and includes details of business tasks. The small business is in the agriculture industry and manufactures products from the main crop. A labour study of another comparable region in the same country \cite{D2016} indicates that similar businesses of the same size as the agriculture business tend to operate on 1.5 FTEs (Full Time Equivalent) staff throughout the year. Seasonal workers are brought in for harvesting. For this analysis, the seasonal workers were not counted towards the employee count as they are contracted for a specific manual task (harvesting) and are likely to have minimal interaction with the business's IT infrastructure.

\subsubsection{Entering Data into the Model}
\label{subsubsec:EnterData}
Using NVivo software, the information for the business operation from the case study was coded to the \gls{sitd} model's classes. When a piece of pertinent information (e.g. job task, person or piece of data) is discussed, it is marked with NVivo code tags of the relevant class together with a unique label (Tag of ``Job Task: Harvest'' is tagged in the text when harvest is mentioned).  Each code tag corresponds to a single object in the object diagram (even if it is referenced/tagged multiple times).  This resulted in 31 codes/objects being generated across Business, Persons, Location, Job Task and Entrepreneurial Characteristics.  Two codes (Destination: Email Host and Product Competition Organiser) were reclassified to Destination System after subsequent public information research. Product Import Data was recoded to Destination System: Email Host from Data Item because the international product importation process relies on email.   Product Competition Organiser was also given a Destination System class as some of the competitions listed in the case study allow for complete online applications.

The objects were initially placed in proximity to other instances of the same class, e.g. job tasks near other job tasks. The case study was then examined again for the relationship between objects.  These relationships are drawn directly in the object diagram.

\subsubsection{Analysing the Result}
\label{sec:AnalysingResult}
\begin{figure*}[h]
{\centering
\includegraphics[width=11cm]{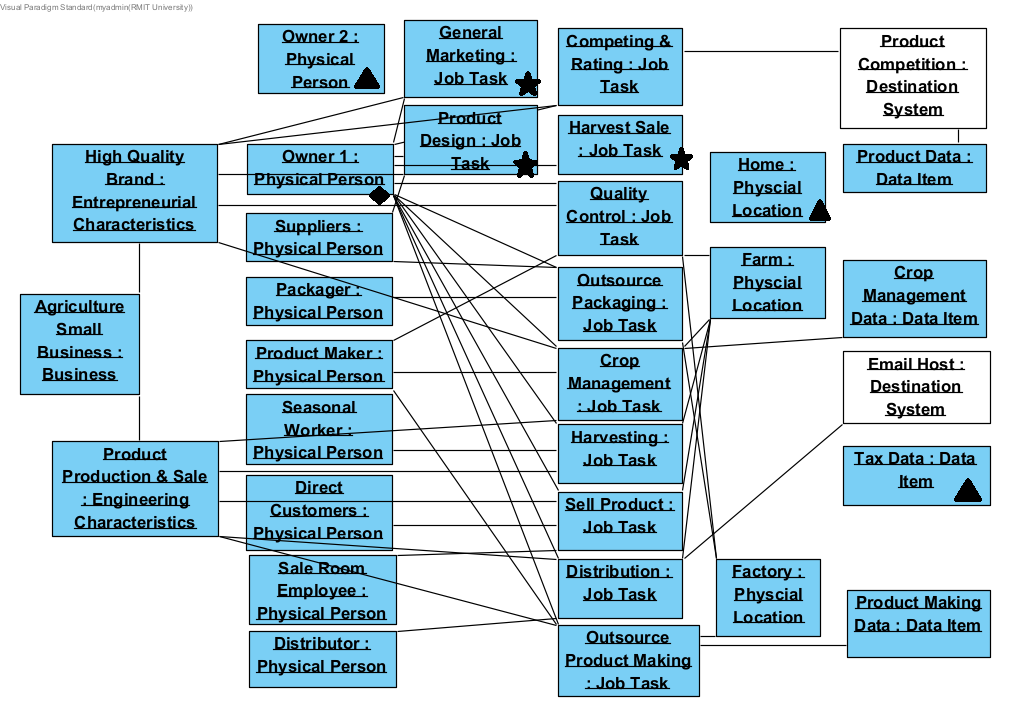}
\caption{\label{fig:ExampleBusinessOverview} Agriculture small business case study result as depicted by the \gls{sitd} model. The reclassified items, Product Competition Organiser and Email Host, are unshaded. Areas of possible investigation and discussion are marked with added symbols. Discussion in section~\ref{sec:AnalysingResult}.}
}
\end{figure*}

The processing of the agriculture small business case study produced Figure~\ref{fig:ExampleBusinessOverview}.

Based on Figure~\ref{fig:ExampleBusinessOverview}, the following areas of hyper or lack of connectivity need discussion from a cyber-security perspective:

\begin{enumerate}
\item \textbf{Critical Point of Failure} - Owner 1 (marked with a diamond \ding{117} in Figure~\ref{fig:ExampleBusinessOverview}) is involved in the majority of the tasks required to keep the small business running.  Any device or system Owner 1 relies on is critical to the business running smoothly, e.g. mobile phone, laptop or cloud services.  Further examination and discussion on additional security controls such as anti-virus and backups, needs to happen to ensure availability of their devices and systems. Owner 1 may need additional training to prevent phishing \cite{Williams2018}.
\item \textbf{Orphaned Components} - There are location (Home), person (Owner 2) and data (Tax Data) items (marked with a triangle \ding{115} in Figure~\ref{fig:ExampleBusinessOverview}) noted in the business, but the information in the case study does not indicate relationships between these and other components. 
\item \textbf{Tasks with No Details} - There are several tasks: Harvest Sale, General Marketing and Product Design, identified (marked with a star \ding{72} in Figure~\ref{fig:ExampleBusinessOverview}), noted in the case study in general, but with no information given regarding the tools/devices needed to perform them.  Most tasks need information, so further exploration is required on whether there are dependencies on IT. 
\end{enumerate}

The lack of information by itself is not treated as a point of concern within the \gls{sitd} model process.  It is an expected by-product of the focus of small business owners, viz to keep business activities running. The purpose of this model is to help obtain the relevant details needed from a cyber-security perspective.

\subsubsection{Task-Based Analysis}
\label{sec:FlowAnalysis}
\begin{figure}[h]
{\centering
\includegraphics[width=\textwidth]{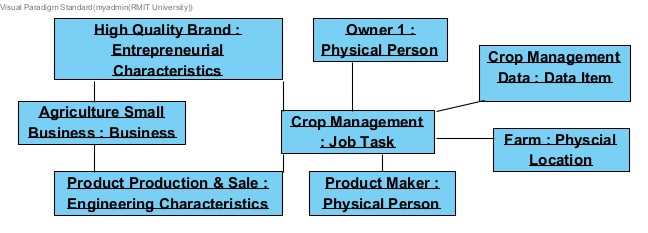}
\caption{\label{fig:ExampleBusinessJob} Elements involved in agricultural small business crop management based on information available in the case study \cite{Osborn2017}. Discussion in section~\ref{sec:FlowAnalysis}.}
}
\end{figure}

The \gls{sitd} model facilitates a job task-oriented approach where each job task is laid out.  From the case study, there is the job task of crop management which covers looking after the crops and managing harvest time. Based on the information available in the case study, we get the object diagram in Figure~\ref{fig:ExampleBusinessJob}.

\begin{figure*}[h]
{\centering
\includegraphics[width=\textwidth]{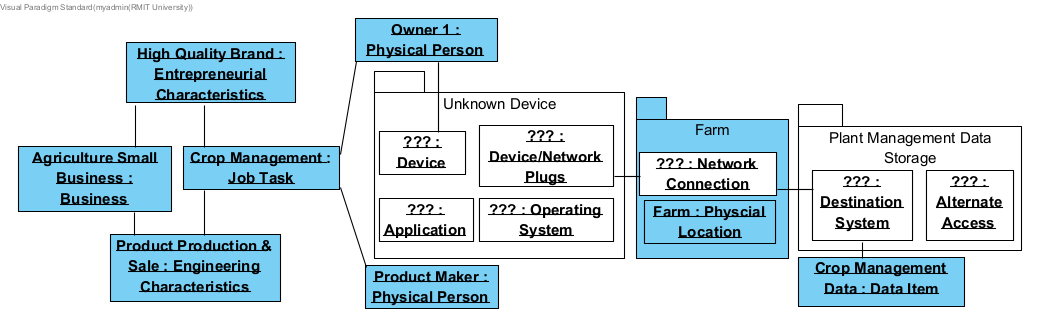}
\caption{\label{fig:ExampleBusinessJobOverlay} Known elements from agricultural small business crop management discovery depicted in the \gls{sitd} model are shaded in blue. Areas for discussion are unshaded. Discussion in section~\ref{sec:FlowAnalysis}.}
}
\end{figure*}

The case study information is inserted into the \gls{sitd} model structure in Figure~\ref{fig:ExampleBusinessJobOverlay}.  Missing information is unshaded.

The unshaded elements indicate missing information that needs further exploration in the cyber-security analysis.  In the case study, crop management decisions are dependent on data like crop ripeness.  Storage, management and access of data is not specified in the case study.  Further information around the devices and data storage needs to be clarified.  By moving through each job task, missing cyber-security context information can be further fleshed out by the cyber-security tool/professional. This information will inform further cyber-security analysis and risk mitigation.

The left side of the diagram in Figure~\ref{fig:ExampleBusinessJobOverlay} clearly links the key reasons for the business to protect crop management data with the core task of crop management.  This relationship information keeps forefront the business value protected by any potential security control. A discussion to protect crop management data is needed as it enables the business to maintain a high-quality brand (by timing the harvest) and supports production. The context helps stakeholders assign the right level of resources and importance.

\subsubsection{Change in Operating Environment}
\label{subsubsec:changelegislativeenvironment}
We now demonstrate the \gls{sitd} model's ability to handle external changes to a small business, with the example of legislative change in the introduction of Australian Goods and Service Tax (GST). GST is a percentage tax that merchants collect on consumer sales \cite{AustralianTaxationOffice2021c}. The collected tax is then passed on to the government in Business Activity Statements (BAS) returns to the Australian Taxation Office (ATO).  Initially, most international sellers were exempt from collection due to the low value of individual orders (such as processed agriculture products in the case study). Subsequently international merchants who achieves a substantial amount of low value sales to Australian customers \cite{AustralianTaxationOffice2021c} were also included. The GST collection requirement applies to our case study business.

To comply with the GST rule using guidance from ATO \cite{AustralianTaxationOffice2021c}, the \gls{sitd} model will be expanded with the following class instances which were not in the case study:
\begin{itemize}
\item ABN: Data Item (Australian Business Number)
\item Australian GST Collected: Data Item
\item Lodge Tax/BAS Return: Job Task
\item Pay GST: Job Task
\item Customs Information: Data Item
\item Customer Invoice: Data Item
\item ATO: Destination System
\end{itemize}

In addition, the following existing instances are modified:
\begin{itemize}
\item Sell Processed Product: Job Task -- Link to additional instances to comply with GST requirements.
\item Production \& Sale: Engineering Characteristics -- Link to the need to lodge additional tax (BAS) returns to the ATO, and the payment of the GST collected using data collected during sales process .
\end{itemize}

\begin{figure*}[h]
{\centering
\includegraphics[width=\textwidth]{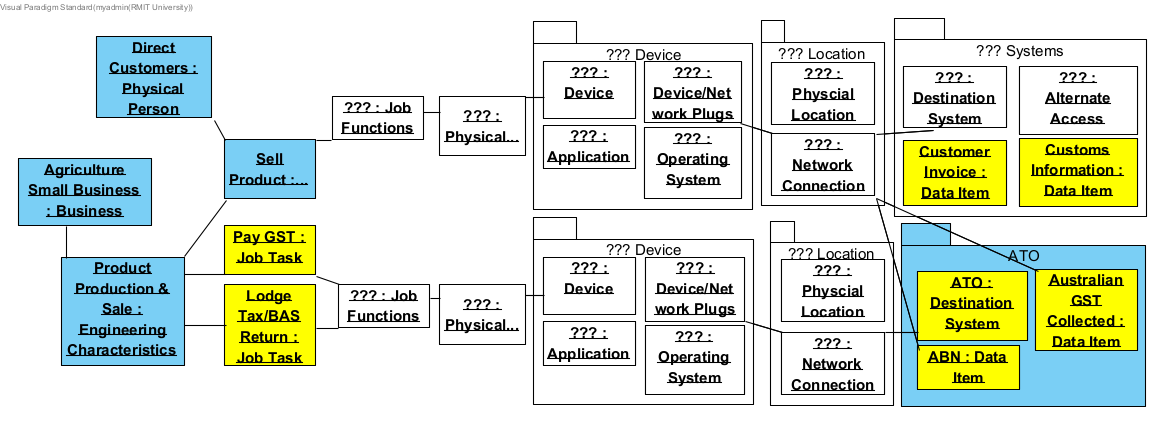}
\caption{\label{fig:GSTChange} GST introduction for Australian customers resulted in additional instances and links in the small business \gls{sitd} model, highlighted in yellow. Missing security component are denoted in non-shaded boxes, denoting areas for further exploration to protect the product production business value. Discussion in section~\ref{subsubsec:changelegislativeenvironment}.}
}
\end{figure*}

The resulting \gls{sitd} model due to external GST change is illustrated in the Figure \ref{fig:GSTChange}, with the added instances highlighted in yellow. Note that the \gls{sitd} structure fundamentally does not change, despite the change in legislative environments. The \gls{sitd} model is designed such that changes to specific environment factors can be handled within the confines of existing \gls{sitd} model structure and items.

Using the \gls{sitd} model structure, the missing instances in Figure \ref{fig:GSTChange} show up as unshaded items. These instances highlight the impacts that the change has on the business in terms of cyber-security.  In this GST example, the \gls{sitd} diagram shows that the business needs to clarify the channels (technology, job function and person) which the required information and tasks entails.  As such, to ensure contiued compliance to the product production and sale, these tools need to be protected proportionate to the business value brought by production.

\subsection{Modelling IT Information}
\label{sec:ITInformation}
\begin{figure*}[h]
{\centering
\includegraphics[width=\textwidth]{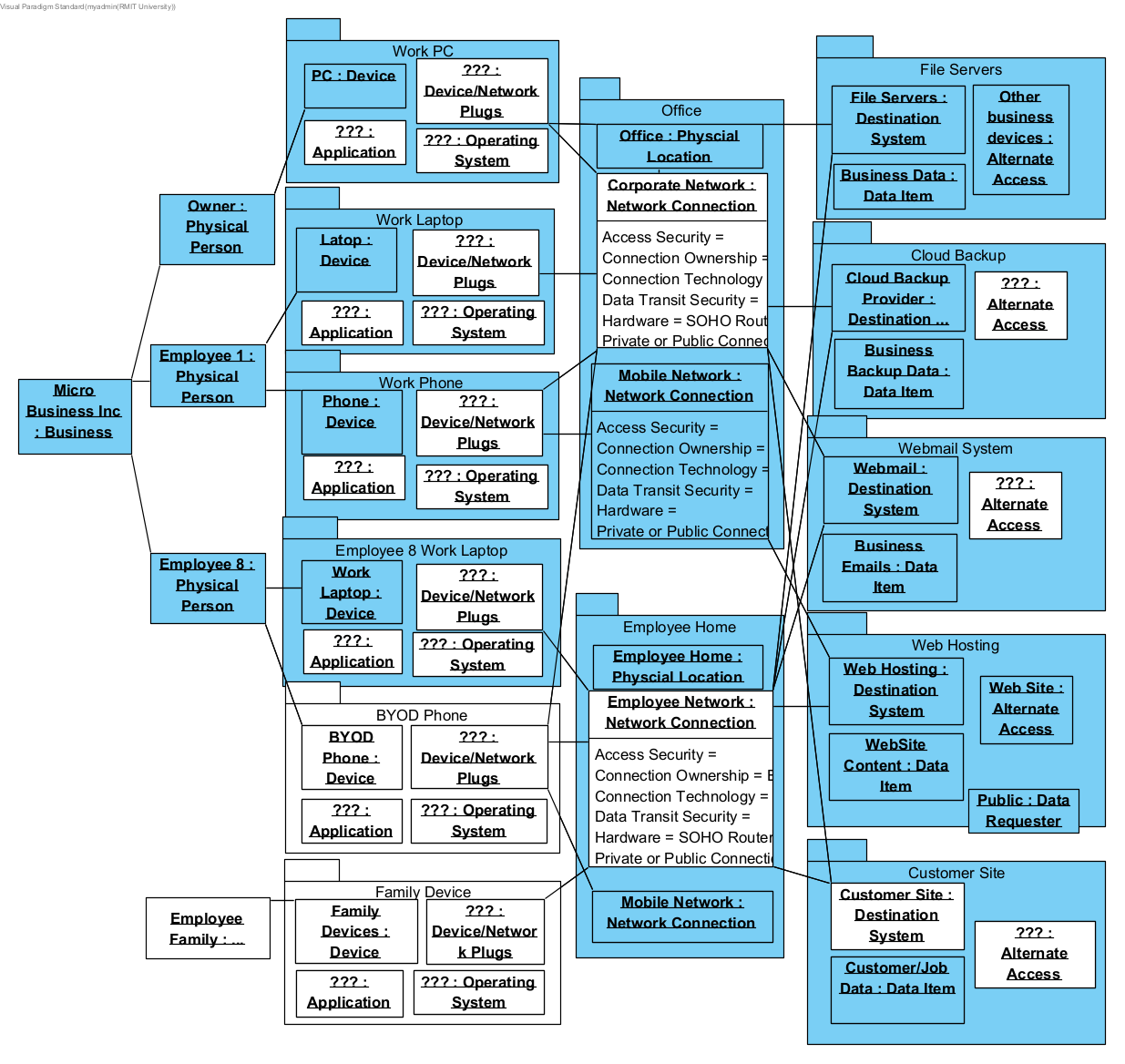}
\caption{\label{fig:OsbornMicroCompany} Model depicting micro company (1-8 employees) architecture. Area highlighted in blue defines components discovered by the case study \cite{Osborn2017}. Any missing information/areas of concern are left unshaded. Discussion in section~\ref{sec:ITInformation}}.
}
\end{figure*}

The case study in Section \ref{subsec:ModelBusinessPoV} was written from a business operations point of view, and contains little technical information.  To emulate, in general, what technical information may be discovered in a small business, we now leverage the small scale IT architectures found in the UK~\cite{Osborn2017}. To be consistent with the previous business operation example of 2 FTE operators, we focus on the micro-companies (1-8 employees) described in the UK study. Figure~\ref{fig:OsbornMicroCompany} gives the overall picture. 

Following the process used in the business operations analysis (Section \ref{subsubsec:EnterData}), components described in the article~\cite{Osborn2017} \footnote{As the business is currently operating, all identifying information of the small business, individuals, specific products and locations are redacted in the public version of this article to protect their privacy, including citations. Original information and citations were provided for review purposes.} were assigned to the corresponding \gls{sitd} classes (e.g. laptop/PC/phone to device class, internet to network connection). Any connected classes, according to the \gls{sitd} model, to the article-identified components that were not discussed in the article were highlighted as missing information in the \gls{sitd} structure between devices and people.

In Figure~\ref{fig:OsbornMicroCompany}, relationships have been assumed to facilitate ease of reading. The relationships between individual employees and their devices, while hard to present in a single page pictorially, can be stored in a data repository easily.

When the typical micro-company case study architecture is mapped onto the structure of the \gls{sitd} model, the following topics are highlighted from a cyber-security perspective:
\begin{enumerate}
\item \textbf{Alternative Access Missing for Cloud Backup and Webmail} -- Cloud backup and webmail are not mentioned in the case study~\cite{Osborn2017}. The business's understanding of the configuration and the ways both systems allow third party access to the business data needs to be evaluated. The discussion needs to be from both technical (e.g. login security, sharing settings) as well as business process (e.g. terms and conditions, privacy policies) perspectives.  The importance of understanding both technical and process aspects is demonstrated in past breaches of misconfigured Amazon S3 buckets \cite{WizCaseCyberResearchTeam2021} and Zoombombing attacks \cite{Greenberg2021}.
\item \textbf{Unknown device applications and operating systems} - A discussion around operating systems in use prompts thoughts around the level of support and processes needed to secure these environments. For example, updating an Android phone and applications is a different process to updating iOS phones. The architecture states that devices are used by both business and personal users. This raises a number of questions. Does the mixed use extend to applications? Are there any applications that are shared with personal use or of untrusted origins? Such cyber hygiene matters are often covered in acceptable use policies in larger enterprises \cite{CenterforInternetSecurity2023} and need to be addressed in small businesses. 
\item \textbf{Sharing the Network} -- The mixed uses of networks are depicted in both the network connection as well as family members sharing that connection. The blending of personal and business use is often not addressed explicitly in cyber-analysis today.  Ideal control conditions explicitly discourage shared network outside the business, which is unrealistic in a small business.  By recognising a blended scenario, discussion can be held around mitigations based on the situation of the small business worker and sharer.
\item \textbf{Mobile Network} -- A phone necessitates a mobile network connection. Most modern phone plans include mobile data. From the case study information, it is unclear whether this network is being used to access the business target systems. Given this network has potential as a backup data connection in the case of an incident, mitigation strategy plan can involve the owner-operator being trained on how to use mobile data to ensure business continuity.
\item \textbf{BYOD Device} -- Workers are permitted to use their own devices and network connection to access business systems. Hence data transit security is another area of concern.  Given that both data transit and devices are not under the small business control, measures may be warranted to secure data depending upon the sensitivity of the data.
\item \textbf{Target Systems Security} - Given the increase in devices connected to corporate systems, the issue of lateral movement during an incident becomes much more prominent.  The breach at a target system (e.g. a cloud provider) can potentially result in a higher impact as attackers can move sideways and onto other devices.  A business needs to understand the nature of the connection to each site/system. This can be strengthened by exploring procedural or technical safeguards that can be applied.  The fallout of supply chain incidents were made clear with past examples such as NotPetya~\cite{Greenberg2018}.
\item \textbf{Customer Site} - Small Office/Home Office (``SOHO'') routers are not typically known for VPN, so in this scenario of connecting to customer site, any work from home employees are most likely connecting via their home network to the customer site.  Clarification around any commercial and security implications around this data transit path is needed.  From a commercial perspective, business needs to understand whether this path contravenes terms of agreement with customer. Alternatively from a technical mitigation perspective, whether VPN can be set up via the existing corporate network routers can influence risk management decisions.
\end{enumerate}

\subsection{Modelling Breach Information}
\label{subsec:applicationinmodellingabreach}


To illustrate the use of the \gls{sitd} model for incident analysis purposes, Maersk's experience from the NotPetya event is modelled within the \gls{sitd} model structure.  The NotPetya incident from Maersk's experience is chosen for the following similarities to small business characteristics:
\begin{itemize}
\item Maersk was ``collateral damage'' \cite{WorldEconomicForum2018} of a wider state attack -- Very few small businesses, especially micro-businesses, have sufficient resources (financial, intellectual property \& data) alone to motivate lone targeted attacks.  Most incidents are likely to be underpinned by an opportunistic element.
\item The attack came from a Maersk's supplier -- Maersk was compelled to use M.E.Doc accounting software to comply with Urkranian accounting norms \cite{Greenberg2018}. This is reflected in small businesses often having very little negotiation power against, or sometimes in the choice of, their IT supplier relationships.
\item Maersk uses IT to support the core business activity of logistics -- Most small businesses are not technical \cite{AustralianBureauofStatistics2020b} and employ IT as a tool to increase productivity. IT is a support function, not their core business.
\end{itemize}

The incident's information was collated using public reports of the incident \cite{Greenberg2018,Greenberg2019}, statements from Maersk's chairman \cite{WorldEconomicForum2018} and accounts of employees involved \cite{Sood2017,Ashton2020}.

\begin{enumerate}
\item A Maersk computer in Odessa (Ukraine) received malicious code through an update from M.E.Doc, a legitimate accounting software. \cite{Greenberg2018}
\item The update contained malicious code, later named NotPetya, which included exploits Mimikatz and EternalBlue. EternalBlue leveraged an SMB version 1 vulnerability to spread the malicious code to systems across the Maersk network. \cite{Sood2017}
\item The malicious code, when encountering a non-patched operating system, encrypted the systems. The encryption effectively wiped the system as NotPetya did not have a built-in decryption capability. \cite{Sood2017}
\item Maersk workers started seeing computers being reset and locked throughout the corporate network \cite{Ashton2020}.  Attempts to limit damage and stop the outbreak were not successful, including physical disconnections of equipment \cite{Greenberg2019}.
\item Workers were sent home as the extent of the computer unavailability became apparent \cite{Greenberg2019}.
\item The assessment and recovery process started on 45,000 PCs, 4,000 servers and 2500 applications \cite{Ashton2020,WorldEconomicForum2018} within Maersk globally. Most of the recoveries, including cleaning and restarting assets, involved labour-intensive processes.
\end{enumerate}

\begin{figure*}[h]
{\centering
\includegraphics[width=\textwidth]{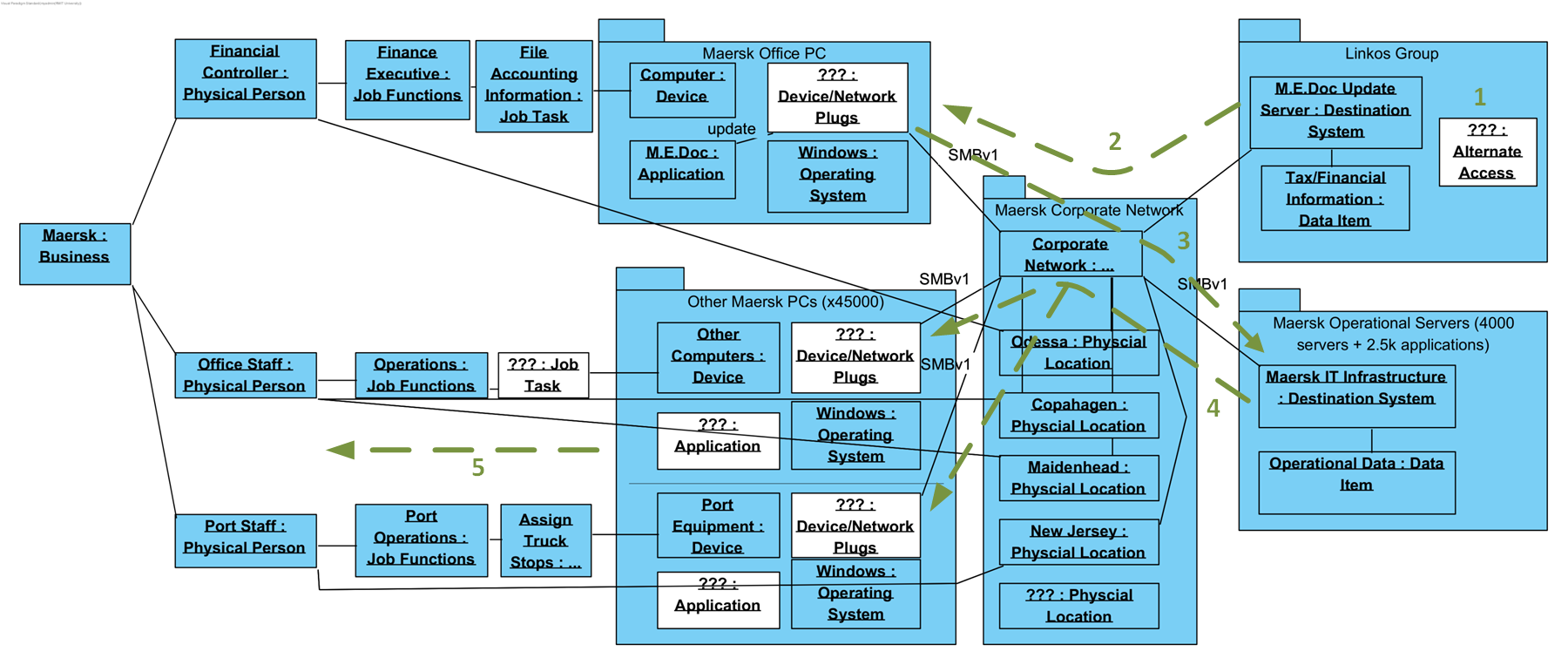}
\caption{\label{fig:MaerskAnalysis}The Maersk NotPetya incident mapped onto the \gls{sitd} model with publicly known information, as discussed in Section \ref{subsec:applicationinmodellingabreach}. Unshaded denotes areas with unknown components.  Dashed arrow lines and numbers in green outline the steps (as listed in Section \ref{subsec:applicationinmodellingabreach}) and infrastructure that was involved in the incident.}
}
\end{figure*}

The \gls{sitd} model illustrates the infrastructure (Figure \ref{fig:MaerskAnalysis}) that enabled the breach, as well as the flow-on human and operational impacts. 

The known information around software trojan from M.E.Docs, the initial point of infection in an office PC and subsequent impact to other Maersk PCs were widely discussed within available sources. These three aspects are recorded in respective Applications and Devices classes. When these elements are plotted on the \gls{sitd} model structure in Figure \ref{fig:MaerskAnalysis}, the roles that the corporate network (Network Connection class) played in disseminating the software are added. This is derived from the fact that NotPetya malware must jump from the originating Office PC to other devices through a network connection. The \gls{sitd} also shows the knock-on impact to operational servers (Destination System class), which in turn led to the dock computer(s) at shipping ports being infected. 

The links between the paralysis of the corporate servers and the shutdown of dock computers, while hinted at and easily derivable by an IT worker, is not always immediately apparent to a non-technical audience. The \gls{sitd} model provides a way to visually `hop' within an often invisible IT architecture, thus laying out the severity and spread of an incident.

\subsubsection{Missing Breach Information}
\label{subsubsec:missingbreachinformation}
In addition, the \gls{sitd} model emphasises the unknown elements, (left unshaded) in-spite of the public information about the incident. The model underscores the insufficient information available about the 
following areas:
\begin{itemize}
\item The corporate network served as the central point of the outbreak. Further investigation from the incident could focus on whether there are any controls allowing ecosystem compartmentalisation. (A lack of network segmentation was raised as an existing issue for the company by an IT employee in a post-incident discussion. \cite{Ashton2020}.)
\item Unknowns around the network connectivity mode of devices (\& associated application and tasks) indicate opportunities for future incident response planning.  As an example, as part of incident response planning, can each physical site identify key (wireless/wired) router(s) that can isolate an entire site? Anecdotally, employees unsuccessfully tried to stop the spread by switching off and/or disconnecting individual workstations in an unplanned manner, not as part of a rehearsed response \cite{Greenberg2018}.
\item The source of entry needs further investigation. How was access (legitimate or not)  to Linkos (maker of M.E.Doc) group's infrastructure granted and/or managed.  While this may not be resolved technically, as the malware was part of a legitimate update, the missing information can drive conversations between Maersk and Linkos as part of the supplier/customer relationship and commercial liability issues.  
\end{itemize}
Examination of the above points can provide potentially valuable information for future cyber-security tightening or setup.  The structure of the \gls{sitd} model highlights gaps or questions that may be overlooked under the pressure of an active incident and post incident reviews. The \gls{sitd} model structure ensures that, even under pressure, a systematic examination of available information can be conducted regardless of completeness.
\section{Discussion}
\label{sec:Discussion}

Section \ref{sec:ApplicationModel} shows how the \gls{sitd} UML model records and maps cyber-security-relevant information for a small business. A business case study, IT architecture case study and breach incident were used to show the \gls{sitd} model recording security-relevant business, and technical and incident details. Due to the socio-technical nature of cyber-security \cite{Schneier2012}, all 3 types of details support the analysis of a business' cyber-security posture.

Table~\ref{tab:PrinciplesCheck} shows, in detail, how the \gls{sitd} model design and usage examples demonstrate the fulfilment of design principles set out in Sections~\ref{sec:TargetUsersBusinesses} and \ref{sec:ModelDeisgnPrinciples}.
\begin{table}[h!]
{\centering
\begin{tabular}{p{5.5cm}|p{1cm}|p{5.5cm}}
\textbf{Design Principles} & \textbf{Met Needs} & \textbf{Section(s) of This Article}\\
\hline
Modelling Small Business (<20 employees) & \checkmark & \ref{subsec:ModelBusinessPoV} \nameref{subsec:ModelBusinessPoV}\\ 
\hline
Usable for Tool Developers & \checkmark & \ref{subsec:advantagesofUML} \nameref{subsec:advantagesofUML}\\
\hline
Business Priorities Focused & \checkmark& \makecell[l]{\ref{sec:BusinessModel} \nameref{sec:BusinessModel} \\ \ref{sec:JobFunctionModel} \nameref{sec:JobFunctionModel} \\ \ref{sec:ThreatModel} \nameref{sec:ThreatModel} \\ \ref{subsec:ModelBusinessPoV} \nameref{subsec:ModelBusinessPoV}}\\
\hline
Capture Intangible Factors \& Relationships & \checkmark & \ref{sec:FlowAnalysis} \nameref{sec:FlowAnalysis}\\
\hline
Allows Incomplete Information & \checkmark & \makecell[l]{\ref{sec:FlowAnalysis} \nameref{sec:FlowAnalysis} \\ \ref{subsubsec:missingbreachinformation} \nameref{subsubsec:missingbreachinformation}}\\
\hline
Language Agnostic & \checkmark & \makecell[l]{\ref{sec:OverallModel} \nameref{sec:OverallModel} \\\ref{sec:FlowAnalysis} \nameref{sec:FlowAnalysis} \\ \ref{subsec:applicationinmodellingabreach} \nameref{subsec:applicationinmodellingabreach}}\\
\hline
Technology Agnostic & \checkmark & \ref{sec:ITInformation} \nameref{sec:ITInformation}\\
\hline
Standards, legislation \& industry Agnostic & \checkmark &\makecell[l]{\ref{subsec:ModelBusinessPoV} \nameref{subsec:ModelBusinessPoV} \\ \ref{subsubsec:changelegislativeenvironment} \nameref{subsubsec:changelegislativeenvironment}}\\
\hline
Enable Cyber-Security Analysis&\checkmark&\makecell[l]{\ref{sec:AnalysingResult} \nameref{sec:AnalysingResult} \\ \ref{sec:FlowAnalysis} \nameref{sec:FlowAnalysis} \\ \ref{sec:ITInformation} \nameref{sec:ITInformation} \\ \ref{subsec:applicationinmodellingabreach} \nameref{subsec:applicationinmodellingabreach}}\\
    \end{tabular}
    \caption{\label{tab:PrinciplesCheck} Design principles (section~\ref{sec:ModelDeisgnPrinciples}) versus proposed model (section~\ref{sec:TheModel}) comparison. The \gls{sitd} model met the principles set out in Sections~ \ref{sec:TargetUsersBusinesses}.}
  }
\end{table}

The modelling of three widely-varying sources of information types demonstrated that the \gls{sitd} model is capable of modelling different types of businesses, situations and information sources.  The \gls{sitd} model's adaptability is useful in cyber-security where small businesses have a variety of different characteristics, ranging from industry nuances to market disruption. Over the long term, this adaptability helps the \gls{sitd} model stay relevant during times of change.

\section{Conclusion}
We proposed the \gls{sitd} UML data model as a way to gather and organise small business cyber-security information. Big standards, whilst flexible, lead to big knowledge requirements and resource commitments, making them difficult to adopt for resource-scarce small businesses. The \gls{sitd} model helps alleviate some barriers faced by small businesses in understanding cyber-security focused security processes and utilising tools currently available. The \gls{sitd} model proposes a new way of working towards a small business cyber-security posture.  

The \gls{sitd} model's analysis of case studies of a micro agricultural business, UK micro-businesses' architecture and NotPetya breach incident shows the capability of the \gls{sitd} model in capturing and organising security-relevant information.  The \gls{sitd} model highlights the value of cyber-security decisions by linking the decisions to the business' operational activities, via \gls{sitd} links between objects. The examples provided above demonstrated the ability of the \gls{sitd} model to model businesses in varied environments, giving structure to an often qualitative, open-ended cyber-security process. 

The \gls{sitd} model's UML foundation gives a ready channel and a structured way for any prospective solution developers (technical or otherwise) to ensure relevant information is captured and organised. Furthermore, UML can readily be accommodated by technologies that allow databases, thus minimising implementation issues and effort.


 The \gls{sitd} model does not seek to replace existing cyber-security standards, but rather fill the existing gaps with respect to small business needs. It is a streamlined way of organising the often informal and piecemeal nature of business information relevant to a cyber-security posture, facilitating the analysis process.  Rather than starting from a technological or risk management perspective, the \gls{sitd} model leads discussion from business-centric perspectives. Structurally, the \gls{sitd} model provides a pathway to connect the business information to IT information.
 
Ultimately, the \gls{sitd} model serves as a pathway towards a more inclusive small business cyber-security process, by taking into account the needs of both cyber-security solution developers and small businesses.

\section*{Acknowledgments}
This research is supported by an Australian Government Research Training Program (RTP) Scholarship. This project was conducted under an Approval from the RMIT Human Ethics Committee (Approval Number: 23928). Authors' Note: The business case study, in section \ref{subsec:ModelBusinessPoV}, citations and identifying information in text and figures have been anonymised to protect the privacy of the small business in publication.  
%
%
%
\bibliographystyle{splncs04}
\bibliography{ModelArticleGeneric}
%





\section*{Appendix - Reading the Model: UML Conventions}
\label{sec:AppendixReadUML}
Our model is highly reliant on UML.  Here is an example to show how to interprete UML class and object diagrams.

Class (drawn as rectangles) defines independent entities or objects. From these classes, associations (drawn as lines) are used to illustrate relationships between classes.  A class can exist on its own but associations must have a beginning and an ending class (which can be the same class). Classes convey structural information only (i.e. the location of the class in relation to other classes). When an instance of a class is defined, it is noted as an object.

Certain characteristics of the class are captured as attributes, primarily text fields within each class.  Enumerations, which restrict the possible values in an attribute, are also defined.  Despite their similarity in appearance to classes, enumerations do not play a role in the structure of the model.

\begin{figure}[h]
{\centering
\includegraphics[width=\textwidth]{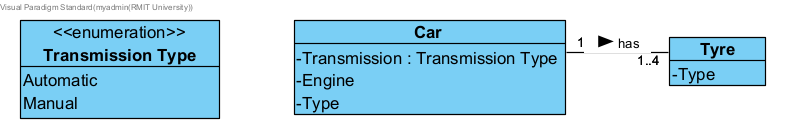}
\caption{\label{fig:ExampleClass} Example class showing the class structure of a car in relation to its tires using UML. The `1..4' notation specifies that a car can have 1 to 4 tyres. Described in section~\ref{sec:AppendixReadUML}.}
}
\end{figure}

Figure~\ref{fig:ExampleClass} illustrates the class diagram for a car for example.

This defines that structurally a car must have a minimum of 1 tyre to a maximum of 4 as a relationship.  These tyres do not need to be the same on the same car (think early bicycles with 2 different size wheels). It also states the attributes of cars and tyres (generally attributes are not mandatory unless otherwise specified). At this stage, it is not instantiated i.e. it does not describe a specific car, but rather a relationship between cars and tyres.

\begin{figure}[h]
{\centering
\includegraphics[width=\textwidth]{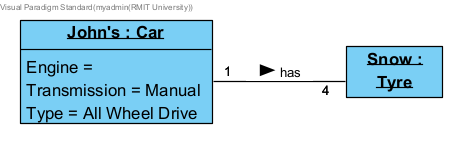}
\caption{\label{fig:ExampleObject} Example object diagram showing how an instance of a car can be represented using UML. Described in section~\ref{sec:AppendixReadUML}.}
}
\end{figure}

In describing individual cars e.g. John's car with four tyres, the object diagram is given in Figure~\ref{fig:ExampleObject}. 

For further information on UML, refer to UML ISO specification documentation \cite{InternationalOrganizationforStandardization2005}.
\end{document}